\documentstyle[emulateapj,psfig]{article}
\def\ie{{\it i.e. \/}}
\def\etal{{\it et al.\/}}
\def\eg{{\it e.g. \/}}

\begin{document}

\submitted{Accepted for publication in ApJ Letters} 

\title{\bf The X-ray Luminosity Function of Nearby Rich and Poor Clusters of 
Galaxies: A Cosmological Probe} 

\author{Michael J. Ledlow$^{1}$, Chris Loken$^{2}$, Jack O. Burns$^2$,
Frazer N. Owen$^3$, \\ and Wolfgang Voges$^4$} 


\affil{$^1$Institute for Astrophysics, University of New Mexico, 
Albuquerque, NM 87131}
\affil{$^2$Office of Research and Dept. of Physics \& Astronomy, University of Missouri \\
 Columbia, MO 65211}
\affil{$^3$National Radio Astronomy Observatory{\footnotemark}, Socorro, NM 87801} 
\affil{$^4$Max Planck-Institut f\"ur Extraterrestrische Physik, Postfach 1603, D-85740,
 \\ Garching bei M\"unchen, Germany}


\footnotetext{The National Radio
Astronomy Observatory is operated by  Associated Universities, Inc., under
contract with the National Science Foundation.}

\begin{abstract} 
  
  In this letter, we present a new determination of the local
  ($z \leq 0.09$) X-ray luminosity function (XLF) using a large,
  statistical sample of 294 Abell clusters and the {\it ROSAT}
  All-Sky-Survey.  Despite the optical selection of this catalog, we
  find excellent agreement with other recent determinations of the
  local XLF.  Given our large sample size, we have reduced errors by
  $\approx$ a factor of two for $L_{X(0.5-2.0~keV)} \geq
  10^{43}h_{50}^{-2}~ergs/sec$.  We combine our data with previous
  work to produce the most tightly constrained local determination of
  the XLF over three orders of magnitude in $L_X$ in order to explore
  possible constraints imposed by the shape of the XLF on cosmological
  models.  A set of currently viable cosmologies is used to construct
  theoretical XLFs assuming L$\propto$M$^p$ and a $\sigma_8 -
  \Omega_0$ constraint (from Viana \& Liddle 1996) based on the local
  X-ray temperature function. We fit these models to our observed XLF and
  verify that the simplest adiabatic, analytic scaling relation (\eg
  Kaiser 1986) disagrees strongly with observations. If we assume that
  clusters can be described by the pre-heated, constant core entropy
  models of Evrard \& Henry (1991) then the observed XLF is consistent
  only with $0.1 < \Omega_0 < 0.4$ if the energy per unit mass in galaxies
  is roughly equal to the gas energy (i.e., if $\beta \sim 1$).

\keywords{cosmology: observations -- galaxies: clusters: general -- X-rays: general}

\end{abstract}


\section{Introduction and Background}

Much of the work on the luminosity distributions of rich clusters has
been motivated by the results of Henry \etal\ (1992) who found
evidence for statistically significant negative evolution in the XLF
(\ie fewer high $L_X$ clusters at higher z) at $z\geq 0.3$ for
$L_{X(0.3-3.5keV)}\geq 5\times 10^{44}h_{50}^{-2}~ergs~sec^{-1}$ from
67 clusters in the {\it Einstein} Extended Medium-Sensitivity Survey
(EMSS).  Recently, Vikhlinin \etal\ (1998) have confirmed the EMSS
result at $z>0.3$ for $L_{X (0.5-2keV)} > 3\times
10^{44}h_{50}^{-2}~ergs/sec$ from a 160 $deg^2$ survey from pointed
{\it ROSAT} fields.  They found a factor of 3-4 decrease
in the number density of these high $L_X$ clusters as compared to a
zero-evolution model.  Several other studies have claimed no
evolution in the XLF out to redshifts as high as $z=0.8$ (Burke \etal\ 
1997; Jones \etal\ 1998; Rosati \etal\ 1998).  However, none of these
studies have sufficiently large search volumes to address evolution in
the XLF at the highest X-ray luminosities and thus do not contradict the
original EMSS result.

Of prime importance in any evolutionary study is an accurate
determination of the {\it local} XLF as a baseline to compare with the
distant cluster XLF.  Until recently, even the local XLF was quite
poorly constrained due to low cluster numbers. The largest local
samples compiled to date are the X-ray Brightest Abell Clusters
(XBACS) (Ebeling \etal\ 1993, 1996) and the Brightest Cluster Sample
(BCS) of Ebeling \etal\ (1997, 1998). The BCS includes 199 X-ray
selected clusters down to $\approx 5\times 10^{42}~\rm ergs~sec^{-1}$
in the $0.1-2.4$keV band out to $z\leq 0.3$.  Consistent with most
previous claims, no evidence was found for evolution in the XLF
within $z\leq 0.2-0.3$ (Ebeling \etal\ 1998).

We have examined a statistically complete sample of 294 Abell rich
clusters within $z\leq 0.09$ using the {\it ROSAT} All-Sky-Survey
(RASS) over the energy band $\rm 0.5-2~keV$ as part of a
multiwavelength study of nearby galaxy clusters.  Unlike most other
studies, our sample is purely optically-selected within the criteria
for inclusion in Abell's catalog.  There is some overlap with both the
BCS and XBACs sample, with the primary differences that we have used
only Abell's northern catalog (Abell 1958), and our X-ray flux-limit
is approximately a factor of 8 lower than the BCS sample.  Our sample
is larger than the BCS while our volume is nearly 30 times smaller.
Given our large sample size, we have reduced statistical errors in the
local XLF for $L_X \geq 10^{43}h_{50}^{-2}~\rm ergs~sec^{-1}$ by up to
a factor of 2 compared to previous work.  Combined with the poor
cluster XLF of Burns \etal\ (1996) (BLL96), we examine the composite
local XLF over more than 3 orders of magnitude in $L_X$ in order to
understand the cosmological constraints imposed by the tight power-law
shape noted in BLL96.

In section 2 we describe the sample, the derivation of the local XLF,
and discuss the limitations imposed by our sample selection.  In
section 3 we compare our new XLF with previous work.  In section 4 we
explore the consequences of the shape of the local XLF with regard to
Press-Schechter analytic predictions of the mass-function and possible
constraints on $\Omega_0$ and $\Lambda$.  We list our conclusions in
section 5.  We adopt $H_0=50~h_{50}~km~sec^{-1}$ and $q_0=0.5$ when
dealing with the observational data.

\section{The Sample and Derivation of the XLF}

Our cluster sample is derived from Abell's Northern catalog, and
includes all Abell clusters in the range $0.016\leq z \leq 0.09$ with
galactic absorption less than 0.1 magnitudes at R-band ($\rm \log N_H
\approx 20.73$).  See Voges \etal\ (1999) and Ledlow \& Owen (1995)
for more details on the sample selection.  The total sample includes
294 Abell clusters. All clusters have measured redshifts and we
include all richness classes in the sample.  We calculate a survey
area of 14,155 $deg^2$ or 34\% of the sky.  Within
our observed volume we find the number density of clusters to be
constant as a function of richness class and redshift suggesting that
our sample is nearly complete and volume-limited within the limits of
Abell's selection criteria.  These findings are consistent with those
of Briel \& Henry (1993) and Mazure \etal\ (1996) with regards to the
completeness of Abell's catalog over this redshift regime.

The X-ray luminosity function was derived from images produced by the
RASS as described in Voges \etal\ (1999). X-ray luminosities were
calculated within a metric aperture of 0.75 $h_{50}^{-1}$ Mpc in
diameter over the energy band 0.5-2 keV assuming a thermal spectrum
with $T=5~keV$. Corrections for missing flux were made according to
the prescription of Briel \& Henry (1993) (using $\beta=2/3$) to
produce a {\it total} $L_X$ for each cluster over our {\it ROSAT}
band. The primary effect of using a different $\beta$ would be 
to shift the total luminosities to higher or lower values (a larger
$\beta$ results in a smaller correction, thus lower total $L_X$),
while not significantly changing the shape or amplitude of the XLF
within the error bars.

Voges \etal\ found a total detection rate of 83\% for this sample of
Abell clusters.  For non-detections, we adopt the $3\sigma$
upper-limits given in their Table 1.  Because of variations in
exposure time (and slight variations in galactic absorption) across
the sky with the RASS, each cluster has a different flux-limit, or
maximum volume to which the cluster could have been detected.  We
follow the prescription of Avni \& Bahcall (1980), and calculate the
observed volume separately for each cluster.  The volume is evaluated
from $z_{min}$=0.016 to the maximum redshift at which the cluster
could have been detected with a $3\sigma$ confidence.  For clusters
with only upper-limits to $L_X$ we set $z_{max}$ equal to the redshift
of the cluster.  The XLF is then found by calculating $dn(L)/dL$ as
the sum over all clusters divided by the maximum search volumes of
each cluster.  Each binned data point is then found by dividing the
above sum by the binwidth ($\Delta L_X$).  For the entire sample, we
find $\langle V/V_{max} \rangle = 0.56 \pm 0.02$.  Error bars on the
data points were calculated assuming Poisson statistics following the
prescription of Rosati \etal\ (1998).

\section{The X-ray Luminosity Function}

In Figure 1 we show the differential XLF for our low-redshift cluster
sample.  Also on this plot are
the measurements of BLL96 derived from 49 poor clusters and the BCS
sample of Ebeling \etal\ (1998).  The steady decline in volume-density
observed in our rich cluster sample for $L_X <
10^{43}h_{50}^{-2}~ergs/sec$ can be understood from the limitations of
Abell's optical selection criteria.  Because $L_X$ varies considerably
for a given optical richness (Voges \etal\ 1999), there are a
significant number of optically poor clusters with $L_X$ in the range
of Richness Class 0 clusters which are not in our sample.  Thus, our
sample is truly volume-limited only for clusters above this cutoff in
$L_X$.  Note, however, that for $L_X > 10^{43}~ergs/sec$, our Abell
cluster sample and the BCS sample are in excellent agreement.  The BCS
also extends to higher $L_X$ because of the larger search volume
($z\leq 0.3$).  Our XLF shown in Figure 1 is also consistent with that of
Edge \etal\ (1990) and Briel \& Henry (1993).

\hbox{~}
\vspace{2.0in}
\centerline{\plotfiddle{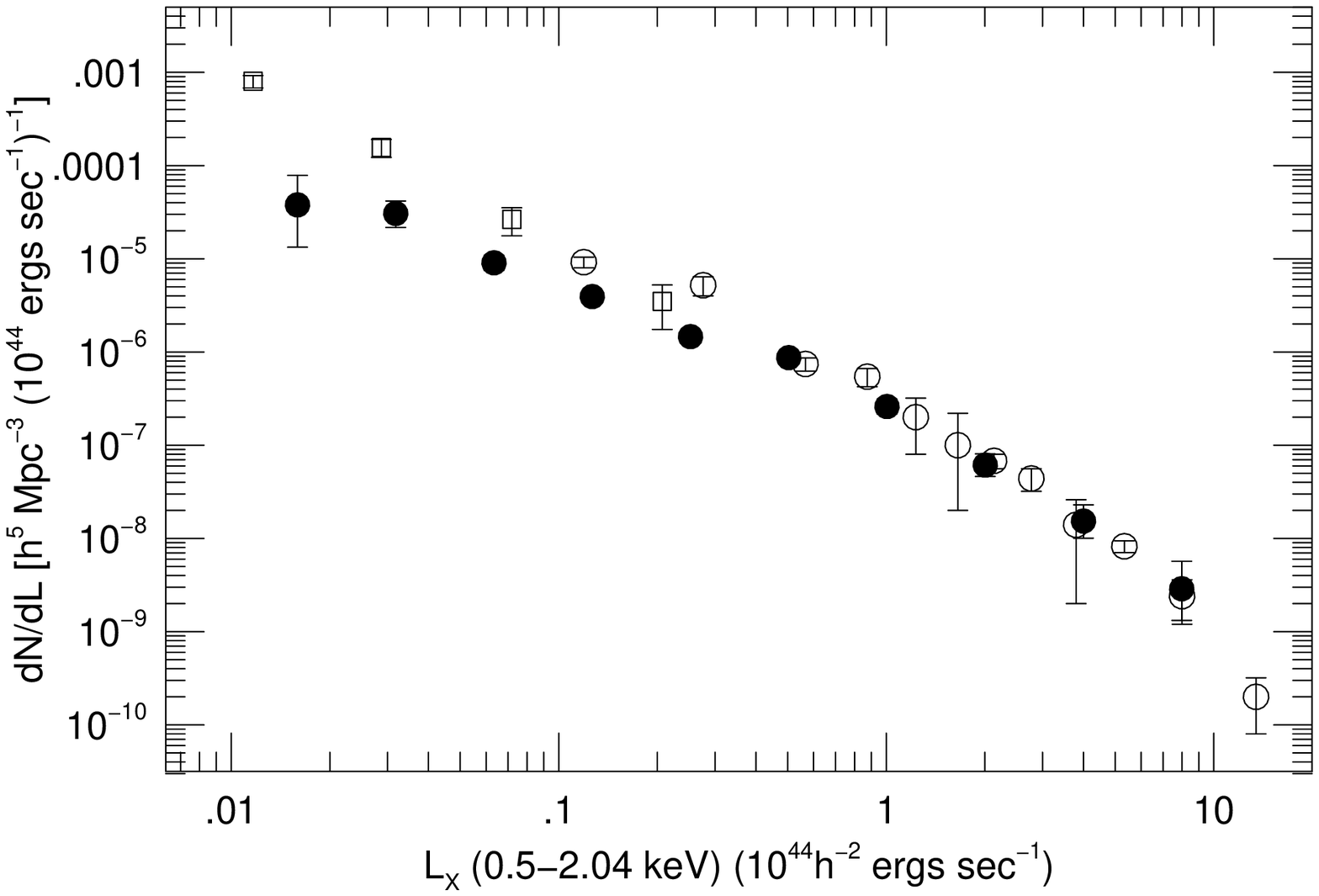}{2.35in}{0.}{45}{45}{-265}{35}} 
\vspace*{-2.0in}
\noindent
{\scriptsize
\hspace*{0.1cm} Fig.~1.\ 
The X-ray luminosity function derived from our
low-redshift ($z\leq 0.09$) Abell cluster sample (solid circles),
the poor-cluster data points from Burns \etal\ (1996) (open
squares), and the XLF from the Brightest-Cluster Sample (BCS) of
Ebeling \etal\ (1998) (open circles).
\addtolength{\baselineskip}{3pt}
}
\smallskip

The local, differential XLF is remarkably well represented by a
power-law over more than three orders of magnitude in $L_X$. The high
luminosity break in the XLF occurs at $> 10^{45}h_{50}^{-2}~ergs/sec$,
and can be seen when we include the highest luminosity point from the
BCS sample.  Using the combined XLF of BLL96 and our new determination
of the local rich-cluster XLF (for $L_X >
10^{43}h_{50}^{-2}~ergs/sec$), we find a power-law fit of the form
$\phi (L)~=~KL_{44}^{-\alpha}$ where $L_{44}$ is the X-ray luminosity
in units of $10^{44}~ergs~sec^{-1}$ and K is in units of
$10^{-7}~Mpc^{-3}~L_{44}^{\alpha - 1}$.  We find best-fit values of
$\alpha=1.83\pm0.04$ and $K=2.35\pm^{0.24}_{0.22}$.  For completeness,
we also fit a Schechter function after including the highest-L$_X$
point from the BCS.  For a fit of the form: ${{dN}\over{dL_X}} = A~
\exp \bigl ( -L_X/L_X^* \bigr )~L_X^{-\alpha}$, we find
A=$(2.93\pm0.14) \times 10^{-7}$ ($Mpc^{-3}~L_{44}^{\alpha-1}$),
$L_{X(\rm 0.5-2keV)}^* =5.49\pm0.39$ ($10^{44}~ergs/sec$), and $\alpha
= 1.77\pm0.01$.  These values are consistent within the errors to the
BCS, the RDCS XLF (Rosati \etal\ 1998) out to z=0.6, and the Southern
SHARC survey (Burke \etal\ 1997) for $0.3<z<0.7$.  Note that these
results do not conflict with the claimed negative evolution in the XLF
observed by Henry \etal\ (1992), and most recently by Vikhlinin \etal\ 
(1998) at the highest luminosities.

As noted by BLL96, the remarkable power-law shape over such a large
range in $L_X$ suggests a continuity in that the bulk X-ray properties
of poor clusters must not be fundamentally different from richer
systems. We explore the consequences of this result in the next
section.

\section{Derivation of the Theoretical XLF}

In order to assess the constraints our local XLF imposes on
cosmological models, we compare it with various analytic
predictions. We proceed by using the Press-Schechter (PS)
formalism (e.g., Press \& Schechter 1974; Bond \etal\ 1991) to
construct theoretical mass functions and then convert these to XLFs
assuming a form for the X-ray mass-to-light ratio (c.f., Evrard \&
Henry 1991; hereafter EH91).

We begin with the set of cosmological models whose parameters are
listed in Table 1. These models form a representative sample of
current views as they include open and flat universes spanning a range
in $\Omega_o$.  For each model, the rms density fluctuation on
8$h^{-1}$Mpc scales ($\sigma_8$) was determined from the $\sigma_8 -
\Omega_o$ relation of Viana \& Liddle (1996) which, in turn, was fixed
by the local number density of 7 keV clusters. The Hubble constant was
chosen to give an age for the Universe of roughly 12.5 Gyrs
(consistent with globular cluster age determinations; e.g., Chaboyer
et al.~1998).  For each model we list the relative contributions of
matter ($\Omega_o$), baryonic matter ($\Omega_b$), and the
cosmological constant ($\Omega_\Lambda$) to the overall energy
density.  Power spectra for all the models were generated using the
code described in Klypin \& Holtzman (1997) and then PS mass functions
(with $\delta_c = 1.3$) were computed at $z=0$.

\begin{table*}
{\scriptsize 
\begin{center}
\centerline{\sc Table 1} 
\vspace{0.1cm}
\centerline{\sc Cosmological Model Parameters} 
\vspace{0.3cm}
\begin{tabular}{lllllllll}
\hline\hline
\noalign{\smallskip} 
{Model}   & {Age$^a$} & 
{$H_o^b$} & {$\Omega_o$} & {$\Omega_b$} & 
{$\Omega_\Lambda$} & {$\sigma_8$} & {p} & {$\beta$} \cr  
\hline 
\noalign{\smallskip} 
OCDM1 & 12.5   & 70 & 0.1 & 0.026& 0.0 & 1.467  & 1.76$^{+.17}_{-.14}$ & 0.66$^{+.07}_{-.05}$ \cr
OCDM2 & 12.7   & 65 & 0.2 & 0.030& 0.0 & 1.162  & 2.20$^{+.19}_{-.17}$ & 0.88$^{+.17}_{-.10}$ \cr
OCDM3 & 12.2   & 65 & 0.3 & 0.030& 0.0 & 1.004  & 2.50$^{+.24}_{-.20}$ & 1.14 $^{+.41}_{-.18}$\cr
OCDM4 & 12.7   & 60 & 0.4 & 0.035& 0.0 & 0.897  & 2.69$^{+.25}_{-.22}$ & 1.47 $^{+.64}_{-.33}$\cr
OCDM5 & 12.3   & 60 & 0.5 & 0.035& 0.0 & 0.817  & 2.86$^{+.32}_{-.21}$ & 1.90 $^{+.65}_{-.5}$ \cr
OCDM6 & 11.6   & 60 & 0.7 & 0.035& 0.0 & 0.702  & 3.22$^{+.29}_{-.29}$ & 4.0 $^{+2.0}_{-2.1}$\cr
$\Lambda$CDM1& 13.1   & 80 & 0.2 & 0.020& 0.8 & 1.478  & 2.14$^{+.18}_{-.15}$ & 0.84$^{+.14}_{-.08}$ \cr
$\Lambda$CDM2& 11.8   & 80 & 0.3 & 0.020& 0.7 & 1.160  & 2.49$^{+.21}_{-.17}$ & 1.17 $^{+.33}_{-.19}$ \cr
$\Lambda$CDM3& 12.4   & 70 & 0.4 & 0.026& 0.6 & 0.901 & 2.77$^{+.27}_{-.22}$ & 1.62 $^{+.63}_{-.37}  $\cr
$\Lambda$CDM4& 12.5   & 65 & 0.5 & 0.030& 0.5 & 0.864 & 2.86$^{+.31}_{-.21}$ & 1.90 $^{+.75}_{-.50} $ \cr
$\Lambda$CDM5& 12.2   & 60 & 0.7 & 0.035& 0.3 & 0.719 & 3.18$^{+.31}_{-.27}$ & 3.5 $^{+2.4}_{-1.6}$ \cr
\noalign{\hrule}
\noalign{\smallskip}
$^a$ current age of universe in Gyrs \cr
$^b$ Hubble constant in units of km/s/Mpc \cr 
\end{tabular}
\end{center}
}
\vspace*{-0.8cm}
\end{table*}

Our PS mass functions can be converted to XLFs by assuming a form for
the mass-luminosity relation and correcting to our bandpass.  We
assume the bolometric X-ray luminosity is related to cluster mass as
$L_{bol}=cM^p$ and will later fit for the parameters $c$ and $p$.
There exist at least two theoretical predictions for the value of the
exponent $p$.  The self-similar model of Kaiser (1986), derived
assuming a power-law initial perturbation spectrum and purely
adiabatic gas physics, predicts $p=4/3$ but it is well known that this
fails to give the correct shape for the XLF (e.g., EH91; see also
below). However, pre-heating of the ICM at an early epoch (possibly by
galaxy formation) results in a different scaling relation and also
resolves several discrepancies between theoretical and observational
results concerning evolution in the XLF (e.g., Evrard 1990, Navarro,
Frenk \& White 1995). For the case of a constant entropy core, EH91
derived a scaling which implies $p = (10 \beta -3)/3 \beta$ where
$\beta = \mu m_p \sigma^2/kT$ is the usual ratio of dark matter to gas
`temperatures'.

We correct our bolometric luminosities to the 0.5-2 keV bandpass by
calculating temperatures and applying a correction appropriate for a
plasma with a metallicity of Z=0.3Z$_\odot$. Specifically, the
temperature corresponding to a given mass can be calculated from the
analytic M-T relation derived from the virial theorem (e.g., Bryan \&
Norman 1998): $kT = \frac{1.39}{\beta}
\left(\frac{M}{10^{15}M_\odot}\right)^{2/3} \Delta_c^{1/3} \; h^{2/3}
~\rm keV$ where $\Delta_c$ is the current density contrast within the
cluster virial radius. The
luminosity in our bandpass is then calculated by applying the usual
bremmstrahlung correction factor as well as a
multiplicative factor to account for the presence of metals (Bryan \&
Norman 1998; eqn.~21).

Using the relation for $L_{bol}$ and the bandpass correction, we
converted our PS mass functions to differential luminosity functions
and made $\chi$-squared fits to a subset of the observational data.
The observational points used in the fits are all four poor cluster
points (BLL96), the five highest luminosity Abell cluster points and
the highest luminosity BCS point from Figure 1.  We first set
$\beta=1$ in the $M-T$ relation and fit for $c$ and $p$.  The fitted
value for $p$ is included in Table 1 and examples of two of the fits
are shown in Figure 2. The dashed curve in Figure 2a is the best fit
when the exponent is kept fixed at the analytic prediction $p=4/3$.
Clearly, the shape of the XLF derived using this prediction is in
gross disagreement with the observed function.  Figure 2b also shows
the importance of the low and/or high-luminosity data points. If only
our five Abell cluster data points are used (dotted line), the fitted
value of $p$ increases by at least 0.2 in all cases (from $p=3.18$ to
$p=3.88$ in this case).  We get virtually identical results if we redo
our fits without the BCS point whereas dropping the poor cluster
points results in slightly greater discrepancies.

\hbox{~}
\vspace{2.0in}
\centerline{\plotfiddle{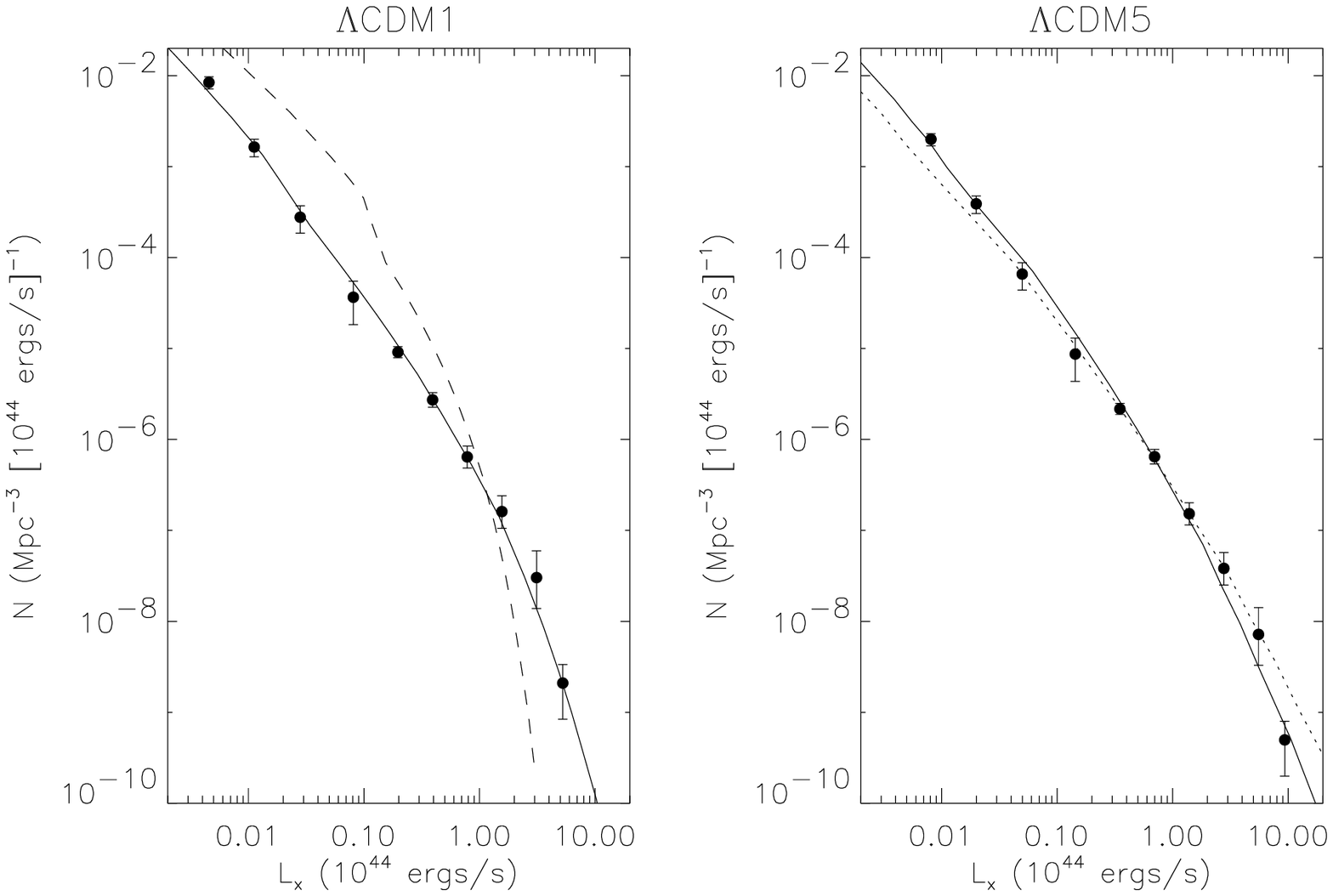}{2.5in}{0.}{45}{45}{-283}{96}} 
\vspace*{-2.0in}
\noindent
{\scriptsize
\hspace*{0.1cm} Fig.~2.\ 
Fits to the observed XLF for two models.  The solid
lines are the best fits to a subset of the observational XLF points
for models $\Lambda$CDM1 (left panel) and $\Lambda$CDM5 (right
panel).  The dashed line in the left panel is the best-fit to the
data assuming $p$=4/3 as predicted by the self-similar analytic
models.  The dotted line in the right panel represents the best-fit
to the data when the BCS and poor cluster points are ignored.
\addtolength{\baselineskip}{3pt}
}
\medskip

If we invoke the constant entropy core model of EH91, then the
exponent in the mass-luminosity relation is actually a function of
$\beta$ ($p = (10 \beta -3)/3 \beta$). In this case, we fit for $c$
and $\beta$ and find the values listed in Table 1.  Interestingly,
only the models with $0.1 < \Omega_0 < 0.4$ are consistent with the
expected value $\beta \sim 1$.  [A recent observational analysis found
$\beta = 0.94 \pm 0.08$ (Lubin \& Bahcall 1993) which is in good
agreement with numerical results (e.g., Eke, Navarro, \& Frenk 1998).]
Thus, if the constant entropy core model of EH91 applies, the
present-day XLF observations suggest a low-density universe but cannot
distinguish between open and flat cases.

\section {Conclusions}

Starting from an optically-selected, statistical sample of Abell
clusters, we have made a new determination of the local XLF to compare
to previous work and more distant cluster samples.  Our cluster sample
is larger than all previous studies, and is contained within a smaller
volume.  For this reason, we have reduced statistical uncertainties in
the local XLF by nearly a factor of two for a limited range in $L_X$
($L_X > 10^{43}~h_{50}^{-2}~ ergs/sec$).  It is only for $L_X <
10^{43}~ergs/sec$ that incompleteness due to the optical selection of
our sample is apparent.  The observed incompleteness is not a failing 
in Abell's catalog, but rather
results from the contribution of poor clusters and groups below
Abell's richness limit.

Combined with the poor-cluster XLF of BLL96, we have examined the
local XLF over nearly three orders of magnitude in $L_X$.  We find
that the local XLF is remarkably well represented by a power-law over
nearly this entire range in $L_X$.  This is significant evidence that
hierarchical formation results in similar cluster properties over a
large range in $L_X$ and mass.  Including the brightest $L_X$ clusters
from the BCS sample which fall above the break in the XLF at $L_X >
10^{45}~h_{50}^{-2}~ergs/sec$, we also performed a Schechter-function
fit which is in good agreement with other recent surveys to much
higher redshift ($z<0.7$), confirming a lack of significant evolution
at these luminosities. 

We have used our new {\it local} XLF to derive a constraint on
$\Omega_0$.  This would appear to contradict a common claim that the
$\sigma_8-\Omega_0$ dengeneracy can be broken only by including the
evolution with redshift ({\it e.g.\/} Bahcall \& Fan 1998). In fact,
PS mass functions for combinations of $\sigma_8$ and $\Omega_0$ that
satisfy a $\sigma_8-\Omega_0$ constraint differ in shape. Borgani
\etal\ (1999) have recently used the shape of the local XLF in order
to constrain $\sigma_8-\Omega_0$ and the shape of the L-T relation.
Including clusters at higher redshift, they concluded that
$\Omega_0=0.4^{+0.3}_{-0.2}$ for open models, and $\Omega_0 \leq 0.6$
for flat models assuming no evolution in the L-T relation; both of
which are consistent with our results. In this work, we have used the
shape of the local XLF, the local number density of 7 keV clusters,
and the PS formalism in order to constrain the cluster M-L relation;
$L_X \propto M^p$.  There is a clear trend for $p$ to increase with
$\Omega_0$ (see also Mathiesen \& Evrard 1998). None of the
theoretical models are consistent with the analytic prediction $p=4/3$
from Kaiser (1986).  If we adopt the constant core-entropy model of
EH91, and the additional constraint that $\beta \approx 1$, the shape
of the local XLF suggests that $0.1 \leq \Omega_0 \leq 0.4$, with no
constraint on $\Lambda$.

\medskip

\noindent 
{\bf Acknowledgements}

This work was supported in part by NASA grants NAG5-6739 and
NAGW-3152, and NSF grant AST-9896039.  We thank Anatoly Klypin and Jon
Holtzman for use of their code and useful discussions.  We also thank
Neta Bahcall and an anonymous referee for helpful suggestions. JOB and
FNO thank MPE for their hospitality during several visits.

\end{document}